\documentstyle[preprint,cite,aps]{revtex}
\newcommand{\be}{\begin{equation}}
\newcommand{\ee}{\end{equation}}
\newcommand{\bea}{\begin{eqnarray}}
\newcommand{\eea}{\end{eqnarray}}
\newcommand{\ba}{\begin{array}}
\newcommand{\ea}{\end{array}}

\newcommand{\bb}{\bibitem}

\begin{document}
\draft

\title{\bf Anisotropic Lifshitz Point at $O(\epsilon_L^2)$}
\author{Luiz C. de Albuquerque$^\ddagger$\footnote{
e-mail:claudio@fma.if.usp.br}
and Marcelo M. Leite$^\star$\footnote{e-mail:leite@fma.if.usp.br}}
\address{}
%\date{March 2001}
\maketitle

\bigskip
\begin{center}
{\small 

$\ddagger$ {\it Faculdade de Tecnologia de S\~ao Paulo -
FATEC/SP-CEETPS-UNESP. Pra\c{c}a Fernando Prestes, 30, 01124-060.
S\~ao Paulo,SP, Brazil}
\bigskip

$\star$ {\it Departamento de F\'\i sica, Instituto
Tecnol\'ogico de Aeron\'autica, Centro T\'ecnico Aeroespacial, 
12228-900. 
S\~ao Jos\'e dos Campos, SP, Brazil}
}
\end{center}
\vspace{0.5cm}

\begin{abstract}
{\it We present the critical exponents $\nu_{L2}$, $\eta_{L2}$ and 
$\gamma_{L}$ for an $m$-axial Lifshitz point at second order in an 
$\epsilon_{L}$ expansion. We introduced a constraint involving the loop
momenta along the $m$-dimensional subspace in order to perform two-
and three-loop integrals. The results are valid in the range 
$0 \leq m < d$. The case $m=0$ corresponds to the usual Ising-like
critical behavior.}
\end{abstract}

\vspace{3cm}

\newpage

Lifshitz multicritical points appear at the confluence of a disordered
phase, a uniformly ordered phase and a modulated
ordered phase \cite{1,Ho}. The spatially modulated phase is 
characterized by a fixed equilibrium wave vector $\vec{k_{0}}$. In 
this phase, $\vec{k_{0}}$ goes continuously to zero as the system 
approaches the Lifshitz point. If this
wavevector has $m$-components, the critical system under consideration
presents an $m$-fold Lifshitz critical behavior. This sort of critical
behavior is present in a variety of real physical systems including 
high-$T_c$ superconductors \cite{2,Ke,Sa}, ferroelectric liquid crystals
\cite{4,Za}, magnetic compounds and alloys \cite{Bece,Bin,Yo}, among others.

In magnetic systems \cite{Se}, the $m$-fold Lifshitz point can be
described by a spin-$\frac{1}{2}$ Ising model on a $d$-dimensional
lattice with nearest-neighbor ferromagnetic interactions as well as
next-nearest-neighbor competing antiferromagnetic couplings along
$m$ directions. This system can be described in a field-theoretic
setting using a modified $\phi^4$ theory with higher order derivative 
terms, which arises as an effect of the competition along the 
$m$-directions. The Lifshitz universality class is
defined by the parameters ($N, d, m$), where $N$ is the number of
components of the order parameter, $d$ is the space dimension of the 
system, and $m$ is the number of competing directions. 

Other examples of field theories containing higher derivative
terms have been investigated in different physical scenarios. In
cosmology, the recently proposed model known as
``$k$-inflation'' describes inflation driven by higher order kinetic
terms for the inflaton scalar field \cite{kessence}.
Another instance which arises in quantum field theory in curved
spacetime, is the quantization of scalar fields with 
high frequency dispersion relation around a classical gravitational
background \cite{corley}. In that case, the higher order term accounts
for deviations from Lorentz invariance. The modified
dispersion relation might arise from an unspecified modification of
the short distance structure of spacetime. A
further generalization of this idea is to modify the large distance 
structure of spacetime allowing higher derivative terms, breaking Lorentz
invariance in the infrared regime as well \cite{carmona}. Thus a better
comprehension of how to calculate arbitrary loop corrections for the
Lifshitz critical behavior should give a clue about the proper
perturbative treatment needed for a general higher order field theory.   

In this work we generalize the method recently developed for the $m=1$
case \cite{leite,albuquerque} to calculate the critical exponents $\eta_{L2}$,
$\nu_{L2}$ and $\gamma_L$ using renormalization group techniques and
the $\epsilon_L$-expansion up to $O(\epsilon_{L}^{2})$, where 
$\epsilon_{L} = 4 + \frac{m}{2} - d$ is the expansion parameter in the
perturbative analysis. We recover the results for the $m=1$ case
obtained in \cite{albuquerque} and show for the first time that the
Lifshitz critical behavior reduces to the Ising-like one for
$m=0$. Thus, the Ising-like universality class $(N, d)$ is contained in a
nontrivial way into the Lifshitz's $(N, d, m)$.

We start with the bare Lagrangian associated with the Lifshitz
critical behavior. It can be written as a modified $\phi^4$ field
theory expressed in the following form:
\begin{equation}\label{1}
L = \frac{1}{2}|\bigtriangledown_{m}^{2} \phi_0\,|^{2} + 
\frac{1}{2}|\bigtriangledown_{(d-m)} \phi_0\,|^{2} + 
\delta_0  \frac{1}{2}|\bigtriangledown_{m} \phi_0\,|^{2} 
+ \frac{1}{2} t_{0}\phi_0^{2} + \frac{1}{4!}\lambda_0\phi_0^{4} .
\end{equation}

The quartic dependence on the momenta along the 
$m$-directions will be manifest in the free propagator. 
Here we will consider the system
at the Lifshitz critical point, defined by the values $\delta_0 = t_0
= 0$. In order to compute the critical exponents,
we need to calculate some Feynman diagrams, namely 
$I_2, I_3, I_4$, and $I_5$ \cite{albuquerque,Amit}. 
Setting $t_0=\delta_0=0$,  

\begin{equation}\label{4}
I_2 =  \int \frac{d^{d-m}q d^{m}k}{[\bigl((k + K^{'})^{2}\bigr)^2 +
(q + P)^{2}] \left( (k^{2})^2 + q^{2}  \right)}\;\;\;.
\end{equation} 
is the one-loop integral contributing to the four-point function,

\begin{equation}\label{5}
I_{3} = \int \frac{d^{d-m}{q_{1}}d^{d-m}q_{2}d^{m}k_{1}d^{m}k_{2}}
{\left( q_{1}^{2} + (k_{1}^{2})^2 \right) 
\left( q_{2}^{2} + (k_{2}^{2})^2 \right) 
[(q_{1} + q_{2} + p)^{2} + \bigl((k_{1} + k_{2} + k')^{2}\bigr)^2]}\;\;,
\end{equation}
is the two-loop \lq\lq sunset'' Feynman diagram of the two-point
function,

\begin{eqnarray}\label{6}
I_{4}\;\; =&& \int \frac{d^{d-m}{q_{1}}d^{d-m}q_{2}d^{m}k_{1}d^{m}k_{2}}
{\left( q_{1}^{2} + (k_{1}^{2})^2 \right) 
\left( (P - q_{1})^{2} + \bigl((K' - k_{1})^{2}\bigr)^2  \right) 
\left( q_{2}^{2} + (k_{2}^{2})^2  \right)}\nonumber\\
&&\qquad\qquad\qquad \times \frac{1}
{(q_{1} - q_{2} + p_{3})^{2} + \bigl((k_{1} - k_{2} + k_{3}')^{2}\bigl)^2}\;\;.
\end{eqnarray}  
is one of the two-loop graphs which will contribute to the 
fixed-point, and

\begin{eqnarray}\label{7}
I_{5}\;\; =&& 
\int \frac{d^{d-m}{q_{1}}d^{d-m}q_{2}d^{d-m}q_{3}d^{m}k_{1}d^{m}k_{2}
d^{m}k_{3}}
{\left( q_{1}^{2} + (k_{1}^{2})^2 \right) 
\left( q_{2}^{2} + (k_{2}^{2})^2 \right) 
\left( q_{3}^{2} + (k_{3}^{2})^2 \right) 
[ (q_{1} + q_{2} - p)^{2} + \bigl((k_{1} + k_{2} - k')^{2}\bigr)^2]} 
\nonumber\\
&&\qquad\qquad\qquad\times
\frac{1} {(q_{1} + q_{3} - p)^{2} + \bigl((k_{1} + k_{3} -
  k')^{2}\bigr)^2} 
\end{eqnarray} 
is the three-loop diagram contributing to the two-point vertex
function. We then choose a special symmetry point in
order to simplify the integrals. 
We set the external momenta at the quartic directions equal to zero,
i.e. $k'=k'_1= k'_2=k'_3=0$, and $K'=k'_1+k'_2$. In
addition, for the four-point vertex, 
the external momenta along the quadratic directions are chosen as 
$ p_{i}. p_{j} = \frac{\kappa^{2}}{4}(4\delta_{ij} -1)$, 
where $p_{1}, p_{2}, p_{3}$ are the independent external momenta, 
and $P = p_{1} + p_{2}$. We fix the momentum scale of the two-point
function through $p^{2} = \kappa^{2} = 1$. We
shall use normalization conditions for the vertex functions along with
dimensional regularization for the calculation of the Feynman diagrams.
 
Let us find out the one-loop integral $I_2$. With our choice of the
symmetry point, and introducing two Schwinger's 
parameters we obtain for $I_2$:

\begin{eqnarray}\label{8}
& & \int \frac{d^{d-m}{q}d^{m}k}{\left( (k^{2})^2 + (q + P)^{2} 
\right) \left((k^{2})^2 + q^{2}  \right)} = \int^{\infty}_{0}\int^{\infty}_{0} 
d\alpha_{1}d\alpha_{2}
\Biggl( \int d^{m}k \,\exp(-(\alpha_{1} + \alpha_{2})(k^{2})^2)
\Biggr) \nonumber \\
& & \qquad\qquad\times\int d^{d-m}q\, \exp(-(\alpha_{1} + \alpha_{2})q^{2}
- 2\alpha_{2}q.P - \alpha_{2}P^{2}) .
\end{eqnarray}
The $\vec{q}$ integral can be performed to give

\begin{eqnarray}\label{9}
&& \int d^{d-m}q \,\exp(-(\alpha_{1} + \alpha_{2})q^{2} - 2\alpha_{2}q.P 
- \alpha_{2}P^{2})\nonumber \\ 
&& \qquad\quad  = \frac{1}{2} S_{d-m} \Gamma(\frac{d-m}{2})
(\alpha_{1} + \alpha_{2})^{- \frac{d-m}{2}} \,\exp(- \frac{\alpha_{1} 
\alpha_{2}P^{2}}{\alpha_{1} + \alpha_{2}})\;\;.
\end{eqnarray}
For the $\vec{k}$ integral we perform the change of variables
$r^2=k_1^2+...+k_m^2$. Now take $z=r^4$. The integral turns out to be:

\begin{equation}\label{10}
\int d^{m}k \,\exp(-(\alpha_{1} + \alpha_{2})(k^{2})^2) =
\Bigl(\frac{1}{4}S_m\Bigr)\Gamma(\frac{m}{4})
 (\alpha_{1} + \alpha_{2})^{- \frac{m}{4}}.
\end{equation}
Using Eqs. (\ref{9}) and (\ref{10}), $I_2$ reads
\begin{eqnarray}\label{11}
&& I_2= \frac{1}{2} S_{d-m}\Bigl(\frac{1}{4}S_m\Bigr)
\Gamma(\frac{d-m}{2})\Gamma(\frac{m}{4})\nonumber\\
&&\quad\times\int^{\infty}_{0}\int^{\infty}_{0} 
d\alpha_{1}d\alpha_{2}\,\exp(- \frac{\alpha_{1} 
\alpha_{2}P^{2}}{\alpha_{1} + \alpha_{2}})\;
(\alpha_{1} + \alpha_{2})^{- \bigl(\frac{d}{2} -\frac{m}{4}\bigr)}.
\end{eqnarray}
The remaining parametric integrals can be solved by a change of
variables followed by a rescaling \cite{abdalla}. The integral is 
proportional to $(P^2)^{-\frac{\epsilon_L}{2}}$.
Now we can set $P^2=\kappa^2=1$. Using the identity 

\begin{equation}\label{13}
\Gamma (a + b x) = \Gamma(a)\,\Bigl[\,1 + b\, x\, \psi(a) + O(x^{2})\,\Bigr],
\end{equation}
where $\psi(z) = \frac{d}{dz} ln \Gamma(z)$, one is able to
perform the $\epsilon_{L}$-expansion when the Gamma functions have non integer
arguments. Altogether, the final result for $I_2$ is:

\begin{equation}\label{14}
I_{2} = \Biggl[\frac{1}{4}S_m S_{d-m}
\Gamma(2-\frac{m}{4})\Gamma(\frac{m}{4})\Biggr]\\
\frac{1}{\epsilon_{L}}\biggl(1 + [i_{2}]_m\,\epsilon_{L}\biggr)\;\;\;,
\end{equation}
\noindent where $[i_{2}]_m =  1+ \frac{1}{2} 
(\psi(1) - \psi(2-\frac{m}{4}))$. From now on, we shall 
absorb the factor inside the brackets in Eq. (\ref{14})
in the definition of the coupling constant \cite{Amit}. Then the
redefined integral is:

\begin{equation}\label{15}
{I}_{2} = \frac{1}{\epsilon_{L}}\biggl(1 + [i_{2}]_m\,\epsilon_{L}\biggr)\;\;.
\end{equation}

Now we shall discuss the two- and three-loop integrals. We introduce a
constraint among the loop momenta in different subdiagrams, along the
quartic directions only \cite{albuquerque}. We wish to highlight this 
approximation here by calculating the integral $I_4$ for $m\neq8$.

After our choice for the external momenta along the quartic
directions, we can write $I_4$ in the following way:

\begin{eqnarray}\label{16}
I_{4}\;\; =&& \int \frac{d^{d-m}{q_{1}}d^{m}k_{1}}
{\left( q_{1}^{2} + (k_{1}^{2})^2 \right) 
\left( (P - q_{1})^{2} +  (k_{1}^{2})^2  \right)} \nonumber\\
&&\nonumber\\
&&\times \int \frac{d^{d-m}q_{2}d^{m}k_{2}}
{\left( q_{2}^{2} + (k_{2}^{2})^2  \right)
[(q_{1} - q_{2} + p_{3})^{2} + ((k_{1} + k_{2})^{2})^2]}\;\;,
\end{eqnarray}  
where we changed variables from $k_2\rightarrow -k_2$. 
We integrate first over the subdiagram $q_2,\,k_2$.
In order to integrate over $\vec{k}_2$ we introduce a constraint
relating $\vec{k}_1$ to $\vec{k}_2$ inside this subdiagram, i.e.,
$\vec{k}_1$ is fixed into the second integral in Eq. (\ref{16}).
If the relation between the two loop momenta is of the form
$\vec{k}_1=-\alpha\, \vec{k}_2$ we can solve the integral in terms of a 
product of Gamma functions and a Hypergeometric function.
The value $\alpha=2$ is singled out when we demand that 
the integral is given in terms of Gamma functions only. This is a 
natural generalization of the $m=1$ case \cite{albuquerque}.
Using Schwinger's parameterization and setting $\vec{k}_1=-2 \vec{k}_2$
in the second integral in Eq. (\ref{16}) we find

\begin{equation}\label{17}
I_{4} = I_{2}\,\int \frac{d^{d-m}{q_{1}}d^{m}k_{1}}
{\left( q_{1}^{2} + (k_{1}^{2})^2 \right) 
\left( (P - q_{1})^{2} +  (k_{1}^{2})^2  \right)} 
\frac{1}{[(q_1+p_3)^2]^{\frac{\epsilon_L}{2}}}\;\;.
\end{equation} 
Performing the integral over $k_1$ we obtain

\begin{equation}\label{18}
I_{4} = I_{2}\,\int_0^1 dz\,\int
\frac{d^{d-m} q_{1}}
{\left( q_{1}^{2} -2z\,P.q_1+zP^2 \right)^{2-\frac{m}{4}}
[(q_1+p_3)^2]^{\frac{\epsilon_L}{2}}}\,\,.
\end{equation} 
Using  a Feynman parameter the integral turns out to be

\begin{eqnarray}\label{19}
I_{4}\,\, = &&\frac{1}{2} I_{2}\,
\biggl(1-\frac{\epsilon_L}{2}\,\psi\bigl(2-\frac{m}{4}\bigr)\biggr)
\frac{\Gamma(\epsilon_L)}{\Gamma\biggl(\frac{\epsilon_L}{2}\biggr)}
\,\int_0^1 dy\, y^{1-\frac{m}{4}}
\,(1-y)^{\frac{1}{2}\epsilon_L-1}\nonumber\\
&&\times \int_0^1 dz
\biggl[ yz(1-yz)P^2+y(1-y)p_3^2-2yz(1-y)p_3.P\biggr]^{-\epsilon_L}.
\end{eqnarray} 
The integral  over $y$ is singular at $y=1$ when
$\epsilon_L=0$. We only need to replace the value $y=1$ inside the
integral over $z$ \cite{albuquerque,Amit}, and integrate over $y$ 
afterwards, obtaining

\begin{equation}\label{20}
I_{4} = \frac{1}{2 \epsilon_{L}^{2}} \Bigl(1 + 
3\;[i_{2}]_m \epsilon_{L}\Bigr).
\end{equation}

The integrals $I_3^\prime$ and $I_5^\prime$ can be solved
using a similar reasoning. They are given by

\begin{equation}\label{21}
I'_{3} = - \frac{1}{8-m}\,\frac{1}{ \epsilon_{L}} \Biggl[1 + 
\biggl([i_{2}]_m + \frac{3}{4-\frac{m}{2}}\biggr)\epsilon_{L}\Biggr],
\end{equation}
 
\begin{equation}\label{22}
I'_{5} = 
- \frac{1}{3\bigl(2-\frac{m}{4}\bigr)}\frac{1}{ \epsilon_{L}^{2}} 
\Biggl[1 + 2\biggl([i_{2}]_m + \frac{1}{2-\frac{m}{4}}
\biggr)\epsilon_{L}\Bigr].
\end{equation}

Note that the leading singularities for $I_2$, $I_4$ are
the same as their analogous integrals in the pure $\phi^4$ theory. 
However, $I_3'$ and $I_5'$ do not have the
same leading singularities for they include a factor of 
$\frac{1}{(2-\frac{m}{4})}$. We then introduce a weight factor for
$I_3'$ and $I_5'$, namely $(1-\frac{m}{8})$, so that they have 
the same leading singularities as in the pure $\phi^4$ theory. This
has the advantage of allowing a smooth transition to the Ising-like
case ($m=0$) from the general Lifshitz anisotropic critical behavior
($m\neq8$) as we shall see next.

The fixed point at two-loop level is given by:
 
\begin{equation}\label{23}
u^{\ast}=\frac{6}{8 + N}\,\epsilon_L\Biggl\{1 + \epsilon_L
\,\Biggl[ \Biggl(\frac{4(5N + 22)}{(8 + N)^{2}} - 1\Biggr )\,[i_{2}]_m -
\frac{(2 + N)}{(8 + N)^{2}}\Biggr]\Biggr\}\;\;.
\end{equation}

With this fixed-point one readily obtains the critical exponents
$\eta_{L2}$ and $\nu_{L2}$:

\begin{eqnarray}\label{24}
&&\eta_{L2}= \frac{1}{2}\epsilon_L^2\,\frac{2 + N}{( 8 + N)^2}\nonumber\\
&&\nonumber\\
&&\qquad +\;\, \epsilon_L^3\,
\frac{(2+N)}{(8 + N)^{2}}\,\Biggl[
\Biggl(\frac{4(5N + 22)}{(8 + N)^{2}} - \frac{1}{2}\Biggr )\,[i_{2}]_m
+\frac{1}{8-m}-\frac{2+N}{(8+N)^2}\Biggr]\;\;,
\end{eqnarray}

\begin{eqnarray}\label{25}
&&\nu_{L2} =\frac{1}{2} + \frac{1}{4}\epsilon_L\,
\frac{ 2 + N }{8 + N}\nonumber\\
&&\nonumber\\
&&\; + \,\frac{1}{8}\frac{(2 + N)} {( 8 + N)^3}
\,\Biggl[2 (14N+40)\,[i_2]_m-2(2+N)+(8+N)(3+N)\Biggr]
\epsilon_L^2\;.
\end{eqnarray}

Using the scaling law 
$\gamma_{L} = \nu_{L2}(2 - \eta_{L2})$, the exponent $\gamma_L$ is

\begin{eqnarray}\label{26}
&&\gamma_{L} =
1 + \frac{1}{2}\epsilon_L\,\frac{2 + N}{8 + N}\nonumber\\
&&\nonumber\\   
&&\qquad  +\;\; \frac{1}{4}\frac{(2 + N)}{(8 + N)^{3}}\,
\Biggl[12+8N+N^2+4\,[i_2]_m\,(20+7N)\Biggr]\,
\epsilon_L^2\,.
\end{eqnarray}

It should be emphasized that $[i_2]_m$ is a universal amount, for the 
dependence on $m$ is encoded in such quantity. The parameter $m$ only
appears in a explicit way at the $O(\epsilon_{L}^{3})$ contribution to
the index $\eta_{L2}$. To our knowledge, the explicit dependence on
$m$ is obtained for the first time at $O(\epsilon_{L}^{3})$ for $\eta_{L2}$.
When setting $(m=1)$ in the formulae above, we recover the exponents
previously reported in reference \cite{albuquerque}. As discussed
there, the two-loop calculation  $(N = 1)$ in three dimensions
yields $\gamma_{L} = 1.45$, in a nice agreement with the numerical
Monte Carlo simulation $\gamma_{L} = 1.4 \pm 0.06$.

The amazing fact obtained using the method outlined here is that the 
critical exponents reduce to the Ising-like ones when $m=0$, for
$\epsilon_{L} \rightarrow \epsilon=4-d$. This means
that the universality class for the $m-$fold Lifshitz point includes 
the Ising-like one for this particular value of $m$ in a nontrivial
way. This provides a unified description of the anisotropic Lifshitz
critical behavior ($m\neq8$, $d\neq m$). This is the
first time that an isotropic behavior ($m=0$) can be recovered from the
most general anisotropic Lifshitz criticality. 

Note that our result for the exponent $\eta_{L2}$ is in agreement with
Mukamel's \cite{Muka} at $O(\epsilon_{L}^{2})$ and is independent on
$m$ at this order. It should not be surprising that the approach fails
to describe the $d=m=8$ case, for the exponent $\eta_{L2}$ is
divergent as can be seen from Eq.(\ref{24}). The approximation made is
not suitable for general isotropic cases $d=m\neq8$ as well, since
there is no preferred directions any longer. Another treatment should be
employed to obtain information along the $m$-dimensional competition
axes, since the symmetry point used here is not suitable to find out
quantities along the competing directions. 

All the results in this work follow from expanding the theory around
its upper critical dimension. The constraint introduced along the
$m$-dimensional subspace is equivalent to expand around the theory
without competition, with $m$ kept fixed. A different field-theoretic method 
has been proposed, based on the expansion around the
number of the $m$ competing directions \cite{sak,Me,Di}. The
$m=2,6$ reported cases are in disagreement with our results. This suggests
that the two approaches are inequivalent.  
 
To conclude, we have calculated the critical exponents associated to
correlations along the $(d-m)$-directions perpendicular to the
competition axes. This was possible because we introduced a constraint
between the quartic loop momenta appearing in different subdiagrams in
higher-loop Feynman graphs. The Lifshitz universality class turns out
to reduce to the Ising-like one for the value $m=0$ at least up to the
loop order considered in this work. In principle, the technique can be
readily generalized to analyze general anisotropic Lifshitz type
critical behavior  with arbitrary powers of the Laplacian at
the competing directions. The study of the tricritical 
Lifshitz points using this formalism is also worthwhile.

\bigskip
{\bf Acknowledgments}

The authors acknowledge partial support from FAPESP, grant numbers
00/03277-3 (LCA), 00/06572-6 (MML), and E. Abdalla for kind hospitality
at the Departamento de F\'\i sica Matem\'atica da Universidade de
S\~ao Paulo.  They would like to thank V. O. Rivelles, 
M. Gomes, and S. R. Salinas for useful discussions.

\end{document}